\documentclass[runningheads]{svmult}

\usepackage{makeidx}   % allows index generation
\usepackage{graphicx}  % standard LaTeX graphics tool
\usepackage{subeqnar}  % subnumbers individual equations
\usepackage{multicol}  % used for the two-column index
\usepackage{physprbb}  % modified textarea for proceedings,
\makeindex             % used for the subject index
\begin{document}
\title*{The First 1-2 Gyrs of Galaxy Formation: Dropout Galaxies from $z\sim 
3-6$}
\toctitle{The First 1-2 Gyrs of Galaxy Formation: Drop out Galaxies from
$z\sim 3-6$} 
%\protect\newline in the Particle Deflection Plane}
% allows explicit linebreak for the table of content
%
%
\titlerunning{Dropout Galaxies from $z\sim 3-6$}
% allows abbreviation of title, if the full title is too long
% to fit in the running head
%
\author{Garth Illingworth and Rychard Bouwens}
\authorrunning{Garth Illingworth \& Rychard Bouwens}
% if there are more than two authors,
% please abbreviate author list for running head
%
%
\institute{UCO/Lick Observatory, Astronomy and Astrophysics
Department,  University of California, Santa Cruz, CA \ \ 95064}

\maketitle              % typesets the title of the contribution

\begin{abstract}

The unique high--resolution wide--field imaging capabilities of HST with
ACS have allowed the characterization of galaxies at redshift 6, less than
1 Gyr from recombination.  The dropout technique, applied to deep ACS $i$,
$z$ images in the RCDS 1252--2927, GOODS and UDF--Parallel fields has
yielded large samples of these objects, allowing a detailed determination
of their properties (e.g., size, color) and meaningful comparisons against
lower redshift dropout samples.  The use of cloning techniques has enabled
us to control for many of the strong selection biases that affect the study
of high redshift populations.  A clear trend of size with redshift has been
identified, and its impact on the luminosity density and star formation
rate can be estimated.  There is a significant, though modest, decrease in
the star formation rate from redshifts $z\sim 2.5$ out through $z\sim 6$.
The latest data also allow for the first robust determination of the
luminosity function at $z\sim 6$.

\end{abstract}

\section{Introduction}

The advent of the HST Advanced Camera, the ACS (Ford et al 2003) has
greatly increased our ability to ``watch galaxies form''.  The sensitivity,
resolution and excellent filter set have provided us with images from which
large samples of high redshift galaxies can be derived.  Of particular
interest are those galaxies with red enough {\it i-z} colors to qualify as
{\it i}-dropouts -- galaxies at redshifts $z \sim 6$.  Such objects have
been the focus of a number of papers over the last year (e.g., Bouwens et
al 2003b, Stanway et al 2003, Yan et al 2003, Dickinson et al 2004).
Spectroscopic confirmation is beginning to appear (e.g., Bunker et al 2003,
Dickinson et al 2004) but is challenging as Weymann et al (1998)
demonstrated with their $z=5.6$ object, which took over 6 hours on Keck.

The current frontier for high redshift objects is at $z\sim 6$ (the
ACS UDF and NICMOS UDF--IR images together will likely extend the
dropout sources to redshifts 7 and beyond, but the samples will be
small).  Rapid changes in the properties of high redshift galaxies
must occur beyond $z\sim6$ and so careful characterization of objects
even those separated by small intervals of time, is a well-justified
goal.  Given this, there is great value in having large samples of
$z\sim3-5$ objects to contrast with $z\sim6$ galaxies.  Though only
$0.2-1.0$ Gyr later in cosmic history, $z\sim3-5$ object samples are
larger and much better characterized, providing key information on
evolutionary changes in high redshft galaxies.

%\section{Figures}
%\begin{figure}[b]

\begin{figure}
\begin{center}
\begin{minipage}[c]{0.5\linewidth}
\includegraphics[width=1.0\textwidth]{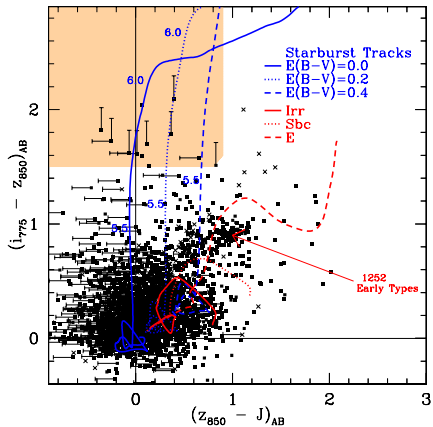}
\end{minipage}\hfill
\begin{minipage}[c]{0.45\linewidth}
\caption[]{Selection of {\it i}-dropouts in the  {\it i-z} {\it z-J}
two--color plane.  This example is for the HST ACS from the RCDS 1252--2927 field
(Rosati et al 1998). The selection limits (particularly the $(i-z)>1.5$ cut --
see Bouwens et al 2003b) returns $z \sim 6$ galaxies with little
contamination (an estimated 11\% contamination rate).}
%\label{eps1}
\end{minipage}
\end{center}
\end{figure}

\section{Fields and Object Selection}

While data from several fields have been used to identify high redshift
dropouts, three fields have stood out for their value for dropout studies
over the last year -- namely the RCDS 1252--2927 field, the GOODS fields,
and the UDF--parallels (UDF--Ps).  All have excellent HST ACS $i_{775}$ and
$z_{850}$ data, while the GOODS and the UDF-Ps fields also have deep
$B_{435}$ and $V_{606}$ data.  The excellent IR data (Lidman et al 2004) in
RCDS 1252--2927 also makes a substantial contribution to the selection of
$i$--dropouts, helping to establish the degree of contamination in the
samples.

The selection of $i$--dropouts is shown in Fig 1 for the RCDS
1252--2927 field (from Bouwens et al 2003b).  The ACS data reaches
typically to $z_{850,AB}\sim 27.3 $ mag ($6\sigma$), while the
ground-based IR data goes impressively deep, down to $J_{AB} = 25.7$
and $Ks_{AB} = 25.0$ mag ($5\sigma$).  The fraction of $z\sim 6$
objects in the IR coverage in RCDS 1252--2927 is impressively small
(0.3\%), only 12 out of $\sim$3000 galaxies.  Even so the
estimated contamination is only about 11\%.  A number of these
candidates have been observed with Keck and the VLT and confirmed to
be at $z\sim 6$.  A total of 23 $z\sim 6$ galaxies are found in four
ACS pointings of the RCDS 1252--2927 field, giving a surface density of
$0.5\pm 0.2$ $i$--dropouts per square arcmin to $z_{AB} = 26.5$ mag.
The objects are very small, though all are resolved, with typical
half-light radii of 0.15$''$ or $\sim 0.9$ kpc.  The $z\sim 6$ objects
reach down to $\sim 0.3 L_{*, z=3}$ (Steidel et al 1999).

Two of the brighter $i$--dropouts from the RCDS 1252--2927 field are
shown in Fig 2, along with their location in the two--color plane, and
SED fits that are used to establish the redshifts.  The ACS $i$ and
$z$ data from the HDF--N also allowed for a search for $i$--dropouts.  
A reassuring result was that the Weymann et al
(1998) object in the HDF--N, spectroscopically verified to be at
$z=5.60$, was very close to meeting our $i$--dropout criterion (its
$i-z=1.2$ color was just a little too blue).  While not a true
$i$--dropout, it suggested that our selection was yielding bona--fide
high redshift objects.  Other spectroscopic results from Bunker et al
(2003) and Dickinson et al (2004), and our own ongoing Keck programs,
have only served to strengthen our confidence in the dropout approach.

%\begin{figure}
%\centerline{\psfig{file=fig1.eps,angle=-90,width=20pc}}
%\caption
%\end{figure} 
\begin{figure} 
\begin{center}
\includegraphics*[width=.95\textwidth]{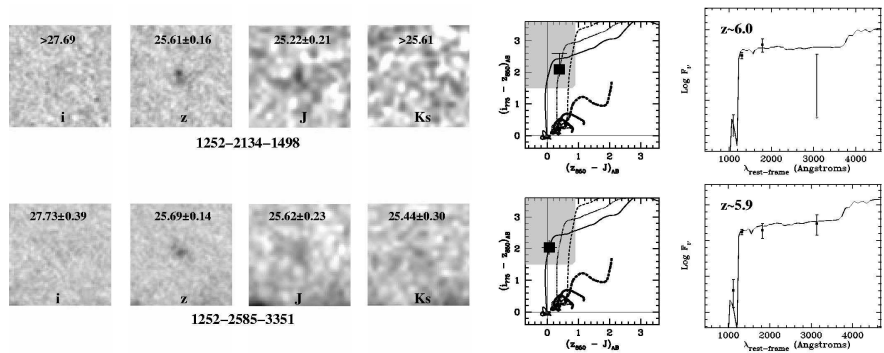}
\end{center}
\caption[]{Images ($3'' \times 3''$) in {\it i, z, J, Ks} of $z\sim 6$ objects,
along with two--color schematics (showing starburst tracks as a function of
redshift for different reddenings - see Fig 1), and starburst galaxy SEDs
($10^8$ Gyr), with the best fit redshift.  The sources are all in RCDS
1252--2927. The magnitudes given for the sources are AB magnitudes.} 
\end{figure}

\begin{figure} \begin{center}
\includegraphics[width=.95\textwidth]{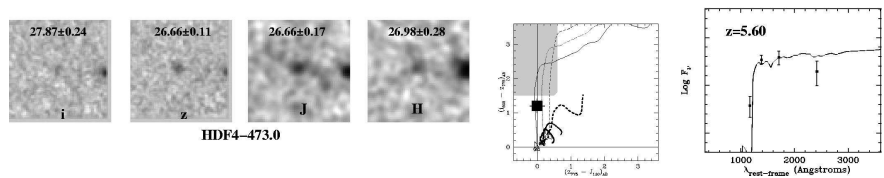}
\end{center}
\caption[]{As in Fig 2, but for the Weymann et al (1998) galaxy in the
HDF--N whose redshift was measured to be $z=5.6$ from over 6 hours of
integration with the LRIS spectrograph on Keck.  The redshift determined
from the photometric data was also $z=5.6$.}
\end{figure}

While the RCDS 1252--2927 field provided a significant sample of
$i$--dropouts (with a good assessment of the contamination from the deep VLT
IR data), the best samples of brighter dropouts come from the two HST ACS
GOODS fields, CDF--S and HDF--N (see Giavalisco et al 2004).  From these
fields, Bouwens et al (2004b) derived a large number of $B$-, $V$- and
$i$--dropouts, augmenting them with a smaller but very useful sample of
$U$--dropouts from the HDF--N and HDF--S fields so that a self--consistent
differential analysis could be applied across a large redshift range,
$z\sim3$ to $z\sim 6$.  Even with relatively conservative selection
criteria, Bouwens et al (2004b) derive 1235 $z\sim 4$ $B$--dropouts, 407
$z\sim5$ $V$--dropouts, and 59 $z\sim 6$ $i$--dropouts.  These samples go as
faint as 0.2, 0.3, 0.5 $L_{*, z=3}$ (using the Steidel et al 1999 value for
$L_{*, z=3}$), respectively, with 10$\sigma$ limiting magnitudes of
27.4 in the $i_{775, AB}$ band and 27.1 in the $z_{850,AB}$ band.

The large samples and wide areal coverage of the GOODS fields are
nicely complemented by the two UDF--parallel fields (UDF--Ps) obtained
in parallel with the deep NICMOS images of the UDF.  These fields have
overlapping ACS images on a 45$''$ grid with 9 orbits
each in B and $V$, 18 orbits in $i$ and 27 orbits in $z$ (as well as 9
orbits with the grism).  They reach impressively faint, to 28.8, 29.0,
28.5 and 27.8 mag ($10\sigma$) in $B_{435}, V_{606}, i_{775}$, and
$z_{850}$ AB-mag, respectively -- or to 0.1--0.2$L_{*,z=3}$.  The UDF
itself will be an impressive addition to these fields, taking the
limits to $<0.1 L_{*, z=3}$.

\begin{figure} 
\begin{center}
\includegraphics[width=.45\textwidth]{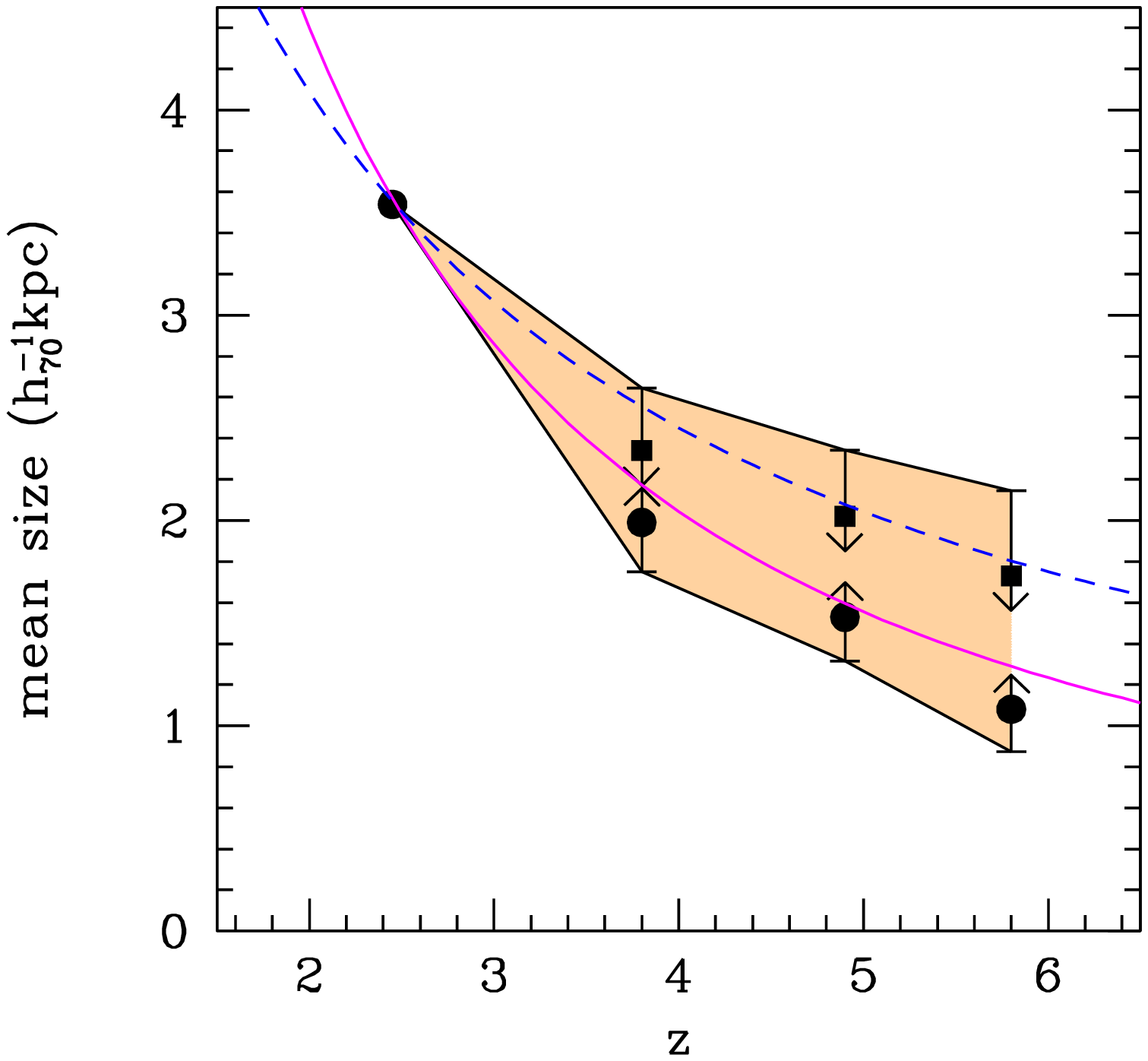}
\includegraphics[width=.45\textwidth]{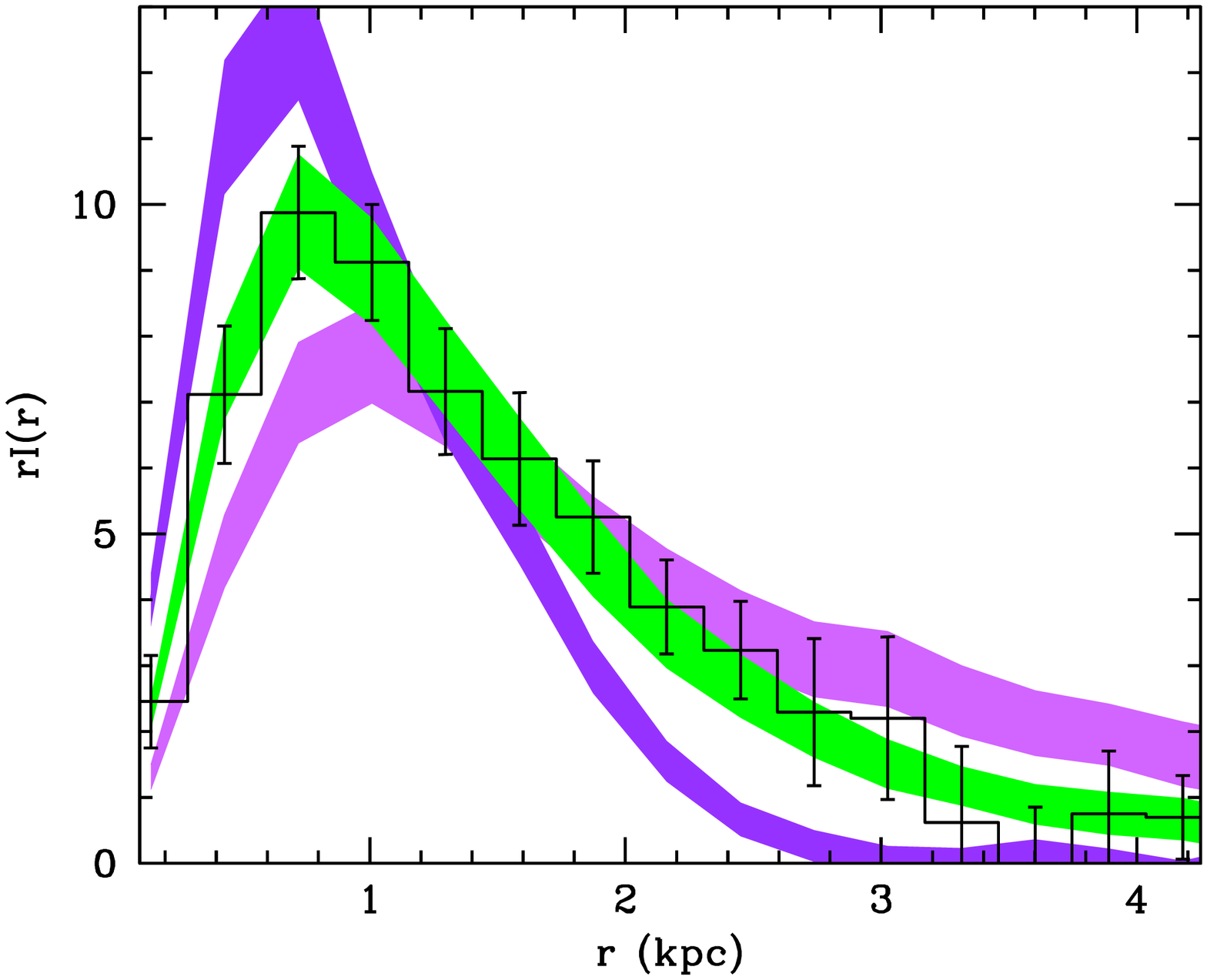}
\end{center}
\caption[]{(Left) Size evolution of $1-2 L_{*,z=3}$ galaxies derived
from composite radial flux profiles for objects with redshifts from
$z\sim 2.5$ to $z\sim 6$ (Bouwens et al 2004b).  The solid black
circles ($1\sigma$ errors) give the observed sizes with redshift, and
should be a lower limit.  The solid black squares are expected to be
upper limits and show the evolution obtained by bootstrapping the
sizes from $z\sim 2.5$, comparing each sample with the one adjacent to
it in redshift (see Bouwens et al 2004b for more detailed
argumentation).  A clear evolution towards smaller size is observed
with redshift, consistent with the scalings predicted from
hierarchical models ($H(z)^{-1}\sim(1+z)^{-3/2}$ for fixed circular
velocity [solid line] and $H(z)^{-2/3}\sim(1+z)^{-1}$ for fixed mass
[dashed line]). (Right) The mean radial flux profile for the 10
brightest $i$--dropouts in the UDF--Ps (histogram) compared with
``cloned'' projections of the HDF--N and HDF--S $U$--dropout sample
scaled in size as $(1+z)^m$, where $m=0$, $-1.5$ and $-3$; $m=-1.5$ is
the best fit (Bouwens et al 2004a). }
\end{figure}

\section{Results}

A major issue with deriving the evolution of galaxy properties at high
redshift is systematic error -- primarily through the many selection
effects that can influence the nature of the samples, even when
derived from very similar datasets.  Of these the $(1+z)^4$ surface
brightness dimming is the dominant effect, but many others affect the
derived samples (e.g., size evolution, color evolution, definition of
selection volumes, data properties as a function of redshift, filter
band, and instrument, etc.).  To treat these effects, we compare our
highest redshift samples with ``cloned'' projections of our lower
redshift samples (e.g., Bouwens et al.\ 1998; Bouwens et al.\ 2003a),
allowing us to contrast intrinsic evolution from changes brought about
by the selection process itself.

\begin{figure} \begin{center}
\end{center}
\includegraphics[width=.5\textwidth]{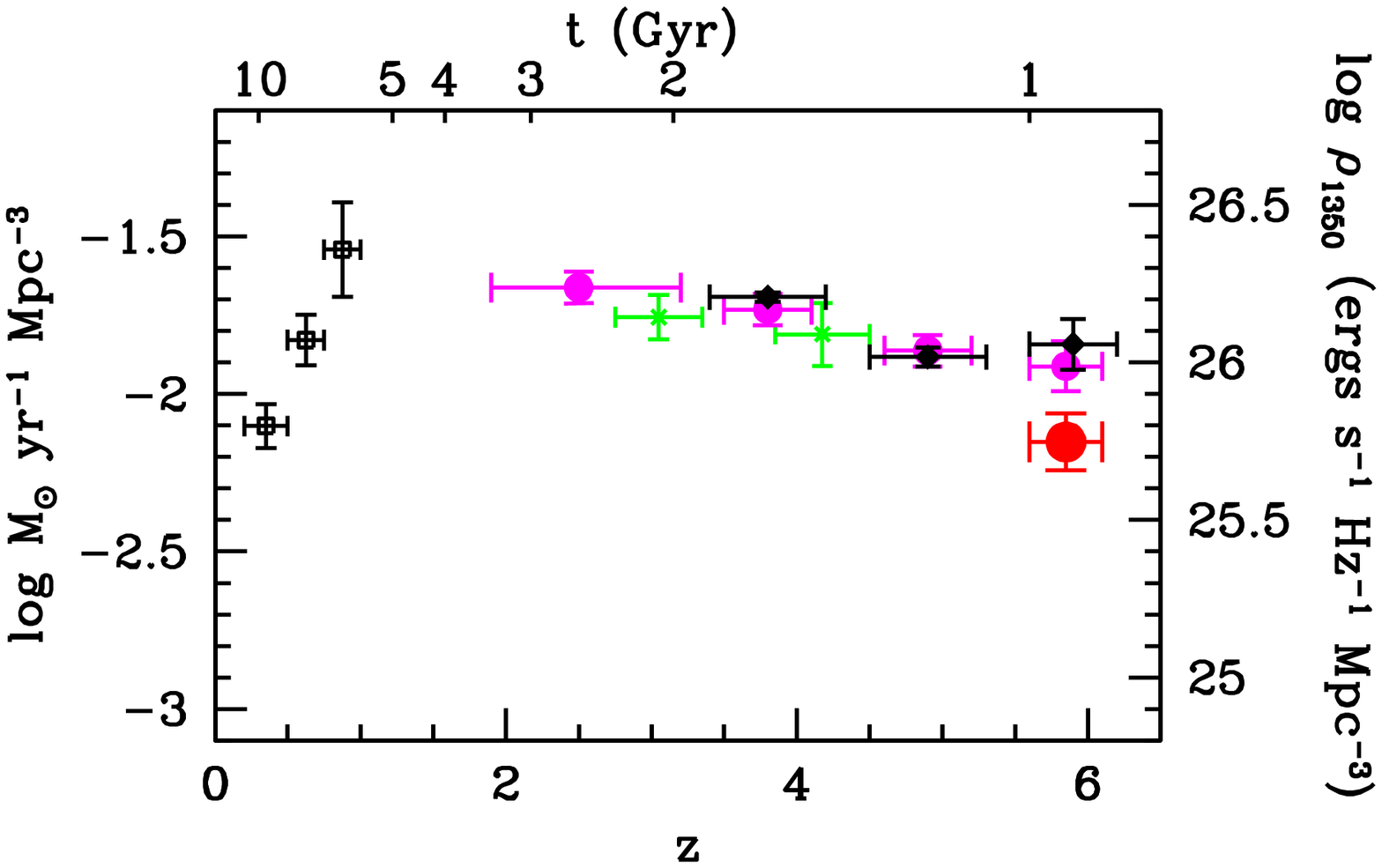}
\includegraphics[width=.5\textwidth]{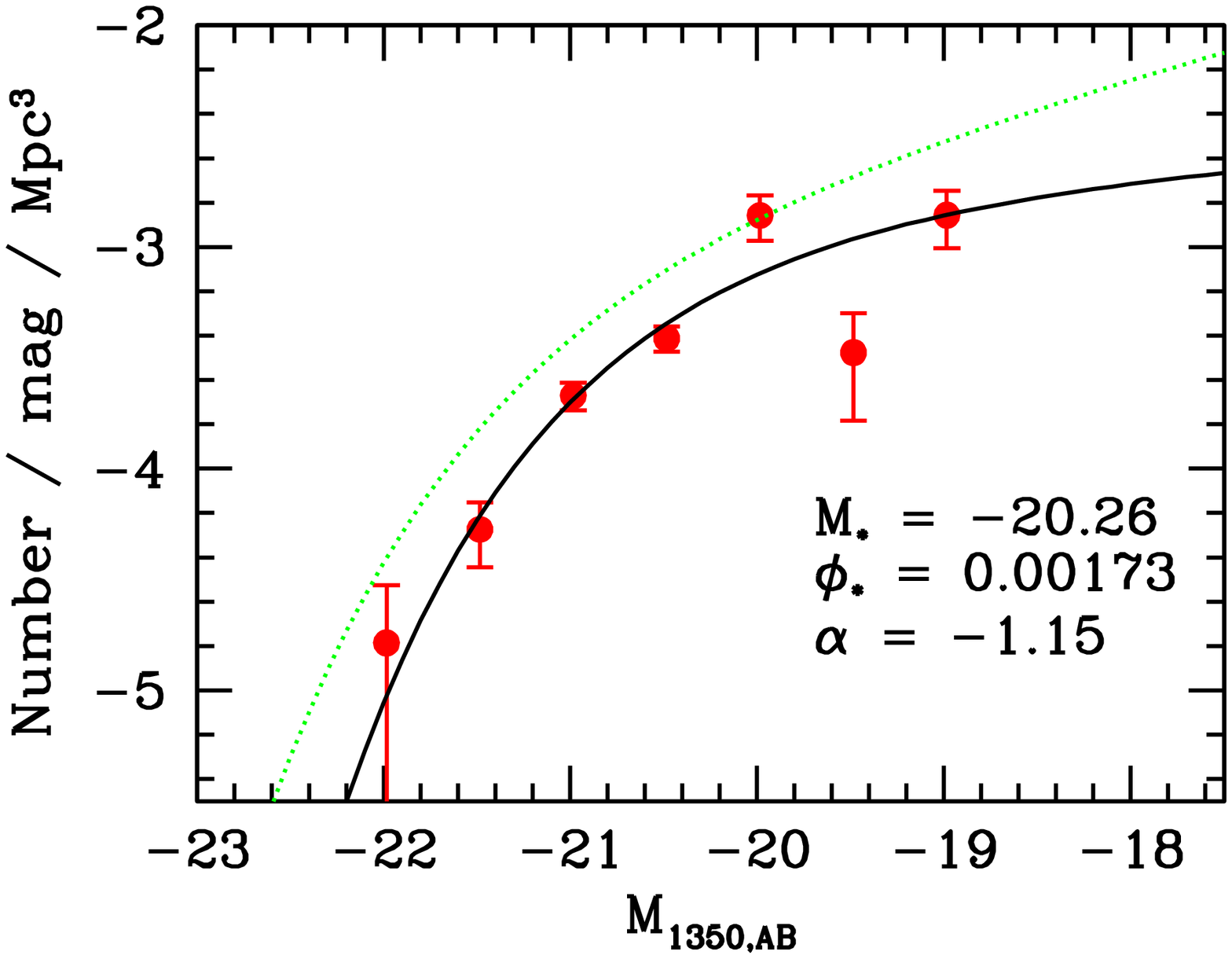}
\caption[]{Star formation rate evolution (dust--free) with
redshift and age (top), integrated down to 0.2$L_{*,z=3}$ (rest--frame UV
continuum luminosity density on the right vertical axis for the high
redshift values).  Note the small $\Delta t$ from $z\sim 6$ to $z\sim 3$.
A Salpeter IMF is used to convert the luminosity density to SFR (see Madau
et al 1998).  (Left)  The four solid circles from $z=2.5$ to $z=6$ are the
values
from the GOODS data (Bouwens et al 2004b).  Other determinations are Lilly
et al (1996 -- open squares), Steidel et
al (1999 -- crosses), Bouwens 
et al (2004a -- solid circles at $z=6$) and Giavalisco et al (2004 -- solid
diamonds). The Thompson et al (2001) values are similar to those shown
here.
The low point at $z=6$ includes
the effect of size evolution on the $z\sim 6$ Bouwens et al (2004a)
value (indicating how significant this effect can be).
(right): The rest frame continuum UV (at 1350 \AA) luminosity function at
$z\sim 6$ from the GOODS field (for $M_{1350,AB} < -19.7$) and the UDF--Ps.
The best fit values for a Schechter luminosity function are shown on the
figure.  The Steidel et al (1999) $z\sim 3$ luminosity function (dotted
line) is also shown.  The Steidel luminosity function had a best fit faint
end slope $\alpha = -1.6$.  Such a  slope is also consistent with our $z\sim
6$ data.
}
\end{figure}

One of the key results is that of size evolution.  This is
demonstrated in Fig 4a for the GOODS fields (see also Ferguson et al
2004).  The deeper UDF--Ps provide stronger and even more conclusive
evidence for this (see Bouwens et al 2004a), as well as indicating
that the best fit appears to be with $(1+z)^{-1.5}$ (Fig 4b).

A major goal of these studies is to extend the constraints on the
luminosity density and the star formation rate with redshift to higher
redshifts $z\sim 6$ and beyond. A related goal is to improve the
constraints at lower redshifts ($z\sim 2-5$).  These new datasets are
proving to be of great value for these two goals.  Fig 5a gives the most
recent estimate of the (dust-free) star formation rate (including the
effect of size evolution) out to $z\sim 6$.

Another major development over the last year is that the HST ACS
data is now of sufficient quality and depth that a luminosity function can
be derived to significantly fainter than $L_*$, as was recently done by
Bouwens et al.\ (2004a) with the GOODS + UDF--Ps data (Fig 5b).  The UDF
will extend this luminosity function one magnitude fainter.

\section{Summary}

There has been a remarkable growth in the number of objects known at high
redshifts ($z\sim 6$) since the HST Advanced Camera came into operation
after servicing mission SM3B. Not only are large numbers of sources being
detected at high redshift, but the development of new techniques for
detecting, characterizing and comparing high redshift objects from
photometric datasets has led to many quantitative results on the nature and
evolution of galaxies in the first $\sim1-3$ Gyrs.

\smallskip
%\acknowledgements

{\bf Acknowledgments} We would like to thank the organizers for an
excellent meeting in a great place. We acknowledge the remarkable advances
that have come about because of HST and its amazing imagers, and regret the
decision to cancel SM4 that will lead to the premature death of HST. We owe
a lot to our team members on the ACS GTO team and the UDF--IR team, and
particularly the PIs, Holland Ford and Rodger Thompson.  Support from NASA
grant NAG5--7697 and  NASA/STScI grant HST--GO--09803.05--A is gratefully
acknowledged.

\def\apj{ApJ.}
\def\apjl{ApJ.}
\def\aap{A\&A}
\def\mnras{MNRAS}

\end{document}